\documentstyle[11pt,paspconf,epsf]{article}

\begin{document}

\title{ Metallicity effects on synthetic Cepheid Period-Luminosity relations.}
\author{I. Musella}
\affil{Osservatorio Astronomico di Capodimonte,  via Moiariello 16, 
I--80131 Napoli, Italy (ilaria@na.astro.it)}

\begin{abstract}
On the basis of new theoretical results (Bono, Marconi \&
Stellingwerf, 1998, hereinafter BMS; Bono, Caputo, Castellani \&
Marconi, 1998, hereinafter BCCM) useful predictions concerning the
Period-Luminosity (PLR) and Period-Luminosity-Color (PLCR) relations
both for optical and infrared magnitudes are presented. It is shown
that, following the dependence of the instability strip on
metallicity, there is a non negligible dependence of the PLRs and
PLCRs on the metallicity of the pulsating stars, mainly for optical
bands. In particular theoretical results predict a dependence of the
PLR on metals which is reversed with respect to current empirical
evaluations (see for instance Gould 1994, Sasselov {\it et al.}  1997,
Kennicutt {\it et al.}  1998, hereinafter K98).  To give a possible
explanation for this discrepancy the typical observational procedures
used to estimate extragalactic distances through Cepheid PLRs are here
tested, with the aim of disentangling, if possible, the reddening and
metallicity effects. To this purpose, synthetic PLRs for different
metallicities were produced and treated as typical observational
samples.
\end{abstract}


\section{Introduction}

The theoretical models computed by BMS are characterized by three different chemical compositions, namely  Y=0.25 - Z=0.004, Y=0.25 - Z=0.008, and Y=0.28 -
Z=0.02, corresponding to the composition of Cepheids belonging to
Small Magellanic Cloud (SMC), Large Magellanic Cloud (LMC) and the Galaxy respectively. 
Following the procedure outlined in K98, I have used the theoretical predictions by BMS and BCCM to construct
simulated Cepheid data sets for the three
different adopted chemical compositions. Cepheid masses have been
derived from a mass distribution given by $dn/dm=m^{-3}$ over the mass
range 5-11M$_\odot$. The corresponding luminosities have been obtained
from the ML relation derived in the assumption of negligible core
overshooting (see BMS and BCCM for details). The location of the predicted instability strip boundaries in the
PL plane is used in order to constrain the position of the simulated
Cepheid samples. At variance with K98 simulation, I did not introduce
artificial foreground reddening to keep, if any, only the
contribution of metallicity.

\section{The simulated PL relations}
In the left plot of Figure 1 the simulated PLRs obtained for the three specified metallicities
and for the 6 bands $BVRIJK$ are shown. As already known,
the dispersion of the PLR decreases moving toward longer
wavelengths. For each PL distribution I have performed both a linear
and a quadratic fit. In each panel the solid line refers to the linear
regression and the dot-dashed line corresponds to the quadratic
one. Moreover, in the case of Z=0.004 and Z=0.02 I have also computed
the least square linear (dashed line) and quadratic (dotted line) solutions
assuming the same dependence on period of the Z=0.008 case and leaving free only the zero-point. In fact, 
usually, in the observational procedure, the extragalactic Cepheid
distance scale is determined by adopting the slope derived from LMC Cepheids. The complete list of the coefficients of these relations will be reported in a future paper (Caputo, Marconi \& Musella, in preparation).

I wish to remark that, in agreement with the discussion in BCCM,
present fits indicate a quite clear overluminosity of metal-poor
Cepheids. 
This trend is  in contrast with several empirical predictions (the
references are quoted in the abstract).  I also note that the middle
and right panels of this plot reveal an intersection between the true
relations for a given metallicity and the corresponding relations
obtained assuming the period dependence of the Z=0.008 case. These
results clearly show that the metallicity effect is critically
dependent on the period range covered by the observational samples.

\begin{figure}
\vbox{
\hbox{
\plotfiddle{musella1a.ps}{8.3truecm}{0}{40}{40}{-410}{-50}
\plotfiddle{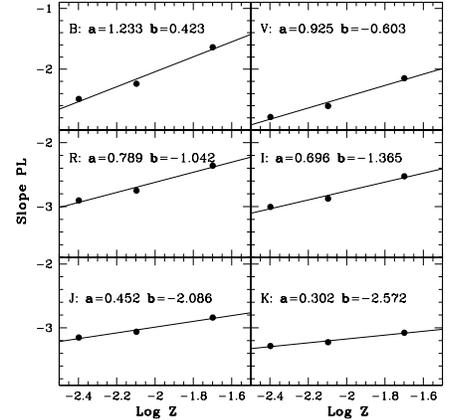}{7truecm}{0}{30}{30}{-565}{-10}
}
}
\vspace{-1.0truecm}
\caption{{\bf Left plot:} Simulated PLRs for the three specified metallicities are plotted. See text for details. {\bf Right plot:} The derived linear PL slopes are plotted versus $\log{Z}$ for the six selected bands. See text for details.}
\vspace{-0.5truecm}
\end{figure}
The right plot in Figure 1 shows the behavior of the linear
PL slopes versus $\log{Z}$ for the six selected bands. The trend of
this dependence is well reproduced by a linear fit ($slope =
a*\log{Z}+b$). It is evident that the slope depends on
metallicity and the effect decreases toward longer wavelengths.
Therefore the assumption of a universal
slope is not an accurate approximation, especially in the optical
bands.

\section{Cardelli law}
One typical observational procedure to determine Cepheid relative
distances is the multiwavelength method described in Freedman {\it et
al.}  (1988). In this procedure, taking advantage of the multicolor
(BVRI with the possible addition of JHK) apparent relative distance
moduli (of the target galaxy with respect to LMC) and assuming that
all the wavelength dependence of these relative moduli is due to the
extinction, it is possible to fit all the data simultaneously with an
interstellar extinction law. In particular, it is assumed that the
reddening law for the LMC and the target galaxy is the same as for the
Galaxy and the law derived by Cardelli {\it et al.} (1989, hereinafter
C89) is used. In this way it is possible to derive the true relative
(to LMC) distance modulus and the total extinction for the Cepheids of
the target galaxy. It is obvious that with this procedure it is
impossible to discriminate between the reddening contribution and the
metallicity one, if any. The HST Key Project, disposing of only two
bands (VI), makes use of a modified version of this method, assuming
that the color excess is equal to the difference between the apparent
distance moduli in the two bands. 

In the present test the zero-point differences (at fixed period
coefficients) can be considered as apparent relative distance
moduli. Since absorption is not included and I am assuming a null
distance for all the Cepheid samples, the  obvious result of the
application of C89 law must be zero for both the true relative
distance modulus and  the absorption coefficient must. Any deviation
from this result can only be ascribed to the metallicity contribution.

The left plot in Figure 2 presents the results of the application of
the C89 law to three different cases: the difference in the
zero-points between the Z=0.004 and the Z=0.008 linear relations, both
with the Z=0.008 slope (upper panel); the difference in the zero-point
between the Z=0.02 and Z=0.008 linear relations, both with the Z=0.008
slope (middle panel); the difference in the zero-point between the
Z=0.02 and Z=0.004 linear relations, both with the Z=0.004 slope
(lower panel). In each panel, the solid line refers to the BVRI C89
fit, whereas the dashed line represents the result of the HST Key
Project procedure. The resulting true distance modulus differences and
absorption coefficients are labeled. I notice that almost all the
contribution of metallicity is included in the absorption coefficient,
while the true relative distance moduli are remarkably close to zero.
In the application of C89 law I have not considered  the NIR bands,
since the apparent relative moduli in J and K are almost identical due
to negligible metallicity effects on these bands. As a result the
application of an extinction law does not make sense.
\begin{figure}
\plotfiddle{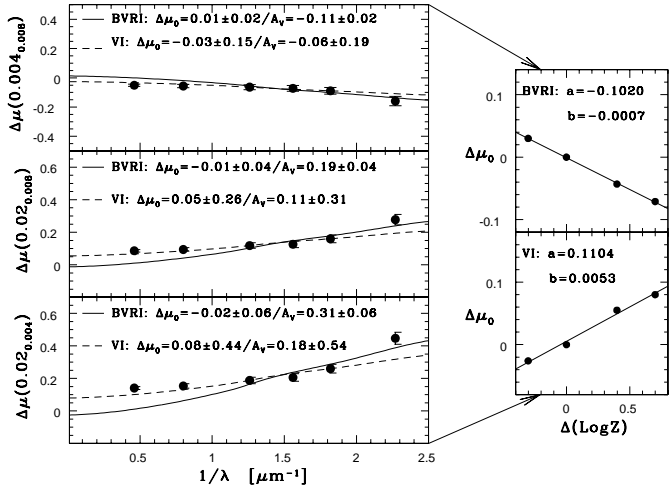}{6.0truecm}{-90}{35}{35}{-140}{200}
\vspace{-0.3truecm}
\caption{{\bf Left plot:} The results of the application of the C89 law to three different cases are presented. See text for details.
{\bf Right plot:} The obtained true relative distance moduli ($\Delta\mu_0$) are plotted  versus $\Delta (\log Z)$ See text for details.
} 
\vspace{-0.5truecm}
\end{figure}
Finally, in the right plot of Figure 2, I report the true relative
distance moduli obtained by the multiwavelength method versus $\Delta
(\log Z)=\log (Z/Z_{ref})$, where $Z_{ref}$ deals with the fixed slope
adopted in the three panels of the left plot of Fig. 2.  An opposite
trend is derived with  the BVRI method and the HST Key Project
approach. In fact, for each $\Delta (\log Z)$, the true relative
moduli  obtained using the two different methods show an opposite
sign, even though in agreement within the errors. I also notice that the VI procedure gives a less accurate value.  As a possible explanation for the
opposite behavior predicted by the two approaches, it is worth
noticing that  the two absorption coefficients are very
different. Indeed, in the HST Key Project, these absorption values are
obtained from color excesses E(V-I)  which are assumed equal to the
differences between the V and I apparent moduli.  Since the
metallicity effects  are differently important  in different bands,
and in particular in V and I, the   quoted procedure may conceal part
of the metallicity contribution,  and even reverse its effect.  Thus,
one could conclude that the reason why the application of the
multiwavelength method and  the VI procedure provides opposite signs
for the metallicity dependence of relative distances has to be
searched in the not proper assumption, in the HST Key project
approach,  that all the difference between V and I apparent relative
distance moduli  is due to absorption effects and that the metallicity
contribution is negligible.

\acknowledgments
I would like to thank Filippina Caputo and Marcella Marconi for their contributions.

\end{document}